\documentclass[pdflatex,sn-mathphys-num]{sn-jnl}

\usepackage[utf8]{inputenc}
\usepackage[T1]{fontenc}
\usepackage{graphicx}%
\usepackage{multirow}%
\usepackage{amsmath,amssymb,amsfonts}%
\usepackage{amsthm}%
\usepackage[title]{appendix}%
\usepackage{xcolor}%
\usepackage{textcomp}%
\usepackage{manyfoot}%
\usepackage{booktabs}%
\usepackage{algorithm}%
\usepackage{algorithmicx}%
\usepackage{algpseudocode}%
\usepackage{listings}%
\usepackage{caption}
\usepackage{subcaption}
\usepackage{xfrac}
\usepackage{nicefrac}
\usepackage{comment}

\theoremstyle{thmstyleone}%
\theoremstyle{thmstyletwo}%

\theoremstyle{thmstylethree}%

\begin{document}

\title[Article Title]{Nonlinear System Identification of Variable-Pitch Propellers Using a Wiener Model}


\author*[1,2]{\fnm{David} \sur{Grasev}}

\author[2,3,4]{\fnm{Miguel} \sur{A. Mendez}}

\affil[1]{\orgdiv{Department of Aviation Technology}, \orgname{University of Defence}, \orgaddress{\street{Kounicova 65}, \city{Brno}, \postcode{66210}, \state{Czech Republic}}}

\affil[2]{\orgdiv{Environmental and Applied Fluid Dynamics Department}, \orgname{von Karman Institute for Fluid Dynamics}, \orgaddress{\city{Rhode-St-Genése}, \postcode{1640}, \state{Belgium}}}

\affil[3]{\orgdiv{Aero-Thermo-Mechanics Laboratory}, \orgname{ \'{E}cole Polytechnique de Bruxelles, Universit{\'e} Libre de Bruxelles}, \orgaddress{\city{Bruxelles}, \postcode{1050}, \state{Belgium}}}

\affil[4]{\orgdiv{Aerospace Engineering Research Group}, \orgname{Universidad Carlos III de Madrid}, \orgaddress{\city{Leganés}, \postcode{28911}, \state{Madrid}}}

\abstract{This work presents the system identification of a variable-pitch propeller (VPP) powertrain, encompassing the full actuation chain from PWM signals to thrust generation, with the aim of developing compact models suitable for real-time digital twinning and control applications. The identification is grounded in experimental data covering both static and dynamic responses of the system. The proposed model takes the form of a Wiener-like architecture, where the PWM inputs are first processed through linear first-order dynamics describing the motor and pitch actuation, and the resulting states are then mapped via a static nonlinear relation to the generated thrust. This structure naturally arises under the assumptions that the electronic actuation operates on a much faster time scale than the mechanical response, and that the contribution of the aerodynamically induced torque is negligible in the tested regime. The resulting parsimonious representation is shown to reproduce the measured dynamics with good accuracy while remaining interpretable and computationally light, thereby providing a practical basis for integration in control-oriented digital twin frameworks.}


\maketitle

\section{Introduction}
\label{sec1}
This paper addresses the system identification of a variable-pitch propeller (VPP) system, encompassing the full actuation chain—from pulse-width modulation (PWM) signals for RPM and pitch commands to thrust generation. The work was conducted within the Re-Twist project (REinforcement TWInning SysTems) at VKI, aimed at developing physics-based digital twins through the integration of system identification, data assimilation, and reinforcement learning. Section~\ref{sec1_1} introduces the Reinforcement Twinning (RT) framework underlying this project. Section~\ref{sec1_2} highlights the opportunities and relevance of VPPs within this context. Section~\ref{sec1_3} reviews existing modeling and identification approaches, and Section~\ref{sec1_4} states the objectives of the present study.

\subsection{The Reinforcement Twinning (RT) framework}
\label{sec1_1}
Reinforcement Twinning (RT) is a recently proposed framework coupling digital twins with control agents through parallel model-based and model-free learning. An adaptive digital twin is continuously updated from real-time data via system identification and data assimilation, while a control agent learns an optimal policy through reinforcement learning or other optimization methods. The twin provides physical consistency and sample-efficient learning, whereas the model-free branch enhances robustness and exploration. Their interaction is bidirectional: the twin acts as a virtual testbed for control strategies, and discrepancies between simulated and real responses drive both model updates and policy refinement. This mutual learning merges the efficiency of model-based and the adaptability of model-free control. RT was demonstrated in wind turbine control, drone trajectory optimization, and cryogenic tank management \cite{Schena2024}. Enhanced variants with stronger corrections were tested for flapping-wing drones \cite{Poletti2025}, and a multi-fidelity Bayesian optimization formulation for attitude control of a two-propeller balance achieved improved learning efficiency \cite{Lecomte2025}. The present work constitutes a first step toward applying RT to VPPs, emphasizing simplified powertrain models and their control performance, thereby laying the groundwork for adaptive digital twins of VPP systems.

\subsection{Opportunities offered by variable-pitch propellers for UAVs}
\label{sec1_2}
In multirotors with fixed-pitch propellers, thrust is controlled solely via motor speed. VPPs introduce blade pitch as a second actuator, enabling thrust modulation through both RPM ($\omega$) and pitch angle ($\beta$). This added degree of freedom allows optimization of agility, efficiency, or endurance \cite{Chang2020}. Early work emphasized agility gains: varying pitch allows faster thrust modulation and reverse thrust, as shown in \cite{Cutler2015}. Yet with advances in brushless motors and speed controllers, modern quadcopters already achieve high agility through RPM control, reducing the incentive for VPP use in lightweight UAVs.

VPPs remain attractive for efficiency improvements. Pitch adjustment maintains favorable blade angles of attack, sustaining higher propulsive efficiency across wider operating ranges \cite{Fresk2013, Fresk2014}. Among pitch–speed combinations yielding equal lift, optimal ones can minimize power consumption and extend endurance \cite{Sheng2016}. Additional benefits include reduced actuator mass in passive concepts \cite{Heinzen2015} and improved fault tolerance under motor failure \cite{Wang2020}. However, modeling VPP powertrains is challenging due to nonlinear coupling between RPM and pitch actuations and unsteady aerodynamic effects, motivating hybrid physics–data-driven modeling approaches.

\subsection{Modeling and identification of variable-pitch propellers}
\label{sec1_3}
Propeller powertrain models traditionally combine electromechanical dynamics with quasi-steady aerodynamic formulations for thrust and torque, derived from Blade Element Momentum Theory (BEMT) \cite{Quintana2018,Podsedkowski2025,Naoki2023,Chang2020} or stationary calibration \cite{Podsedkowski2020,Simmons2022,Porter2016,Kulkarni2024,Heinzen2015}. Linearized electromechanical models, often first-order, are widely used in control and identification studies \cite{Fresk2013,Fresk2014,Chang2020,Quintana2018,Sheng2016,Naoki2023,Michel2022,Sun2018}. 

For VPPs, thrust depends nonlinearly on both RPM and pitch. Sudden pitch changes cause torque variations and direct thrust feed-through \cite{Cutler2015}, which single-state models cannot capture. Recent studies incorporated thrust feedback into actuator models to better represent these couplings \cite{Wu2024}. Some authors relaxed quasi-steady assumptions by introducing inflow lags \cite{Asper2023} or adaptive estimation of aerodynamic coefficients during flight \cite{Ahsun2015}. Although full unsteady coupling remains underexplored, experimental studies confirm measurable unsteady effects in turbulent conditions \cite{Oo2023}. Nonetheless, quasi-steady models remain effective for hover and moderate maneuvers when properly calibrated \cite{Podsedkowski2020,Quintana2018,Simmons2022,Wu2024,Gauthier2023}, even though they fail in fast transients where induced circulation and vortex shedding dominate.

\subsection{Objectives of the present study}
\label{sec1_4}
This preliminary study aimed to establish an experimental test bench enabling thrust and RPM measurements under controlled speed and pitch actuation, as well as varying loads. The setup, initially developed within \cite{Colonval2025}’s STP program, required refinement for systematic identification and control tests. The second goal was to evaluate the feasibility of representing the entire powertrain by an adaptive Wiener-type model, derived from first principles and identified using an adjoint-based procedure. Finally, the study includes initial model-based control experiments for RPM and pitch actuation with step thrust commands. The long-term objective is to build on these results toward an efficient, accurate framework for modeling and control of VPP powertrains in realistic UAV applications.

\section{Dynamic Modeling of the Variable-Pitch Propeller}
\label{sec2}
We first present a general nonlinear model (Section~\ref{sec2_1}), then reduce it under simplifying assumptions to the proposed Wiener model (Section~\ref{sec2_2}).

\subsection{A general electromechanical–aerodynamic model}\label{sec2_1}
In hover (zero free-stream), simple aerodynamic models relate thrust $T(t)$ to rotational speed $\omega(t)$ and blade pitch $\beta(t)$ as
\begin{equation}
\label{Thrust}
T(t) = \tfrac{1}{2}\,\rho\,A\,\omega(t)^2\,R^2\,C_T\bigl(\lambda(t),\,\beta(t)\bigr),
\end{equation}
where $\rho$ is air density, $A=\pi R^2$ the disk area, and $C_T$ the non-dimensional thrust coefficient, a static map of inflow ratio $\lambda=v_i/(\omega R)$ (with induced velocity $v_i$) and pitch $\beta(t)$.

In maneuvers, thrust transients are modeled by assigning dynamics to $\omega(t)$, $\beta(t)$, and $\lambda(t)$ while keeping $C_T(\lambda,\beta)$ static. Equation~\eqref{Thrust} then holds at all times; transients arise from inflow and actuation dynamics. Such models capture wake–induced lags but neglect explicitly unsteady effects (e.g., hysteresis, delayed stall, vortex–induced loads).

The inflow ratio is often modeled as a first-order system \cite{Asper2023}:
\begin{equation}
\label{eq:lambda_dyn}
\frac{\text{d}\lambda(t)}{\text{d}t} = \frac{1}{\tau_\lambda}\,\Big(\lambda_{qs}\!\bigl(\beta(t)\bigr) - \lambda(t)\Big),
\end{equation}
where $\lambda_{qs}$ is the quasi–steady inflow from momentum/blade-element consistency condition and $\tau_\lambda$ is a wake time constant (on the order of disk transit time).

Rotor speed follows the torque balance coupled to the motor circuit:
\begin{align}
    J\frac{\text{d}\omega}{\text{d}t} &= k_Q\,I_m  - k_\omega\,\omega - Q_a(\lambda,\beta), \label{eq:rotordyn}\\[6pt]
    L_m\frac{\text{d}I_m}{\text{d}t} &= u_\omega V_{\text{in}} - R_m I_m - k_i\,\omega, \label{eq:motorelec}
\end{align}
where $J$ is rotor–propeller inertia, $I_m$ the motor current, $k_Q$ the torque constant, $k_\omega$ viscous losses, and $Q_a=\tfrac{1}{2}\rho A(\omega R)^2 R\,C_Q(\lambda,\beta)$ the aerodynamic torque (with static map $C_Q$). In \eqref{eq:motorelec}, $L_m,R_m$ are motor inductance/resistance, $V_{\text{in}}$ the supply, $u_\omega\!\in[0,1]$ the PWM duty cycle, and $k_i$ the back–EMF constant. Equations~\eqref{eq:rotordyn}–\eqref{eq:motorelec} describe current–torque–speed coupling; they can be combined into a second-order nonlinear equation for $\omega(t)$.

Pitch dynamics similarly couple actuator mechanics and circuit:
\begin{align}
    J_a\frac{\text{d}^2\beta}{\text{d}t^2} + D_a\frac{\text{d}\beta}{\text{d}t} &= k_a I_a, \label{eq:pitchmech}\\[6pt]
    L_a\frac{\text{d}I_a}{\text{d}t} &= u_\beta V_{\text{in}} - R_a I_a - k_{ia}\frac{\text{d}\beta}{\text{d}t}, \label{eq:pitchelec}
\end{align}
with $J_a,D_a$ the actuator inertia/damping, $I_a$ the actuator current, $k_a$ the torque constant, $L_a,R_a$ the actuator inductance/resistance, $u_\beta\!\in[0,1]$ the duty cycle, and $k_{ia}$ the back–EMF constant. Equations~\eqref{eq:pitchmech}–\eqref{eq:pitchelec} link electrical input, current, and pitch motion.

The full model \eqref{Thrust}–\eqref{eq:pitchelec} thus comprises states $\omega(t)$, $\beta(t)$, $\lambda(t)$ and hidden electrical states $I_m(t)$, $I_a(t)$, driven by inputs $u_\omega,u_\beta$. Even with quasi–steady $C_T(\lambda,\beta)$ and $C_Q(\lambda,\beta)$, the system is nonlinear. Thirteen parameters must be identified ($\tau_\lambda$, $J$, $J_a$, $D_a$, $k_Q$, $k_\omega$, $k_a$, $R_m$, $R_a$, $L_m$, $L_a$, $k_i$, $k_{ia}$), in addition to the static maps $C_T$ and $C_Q$.

\subsection{The proposed Wiener model}\label{sec2_2}
The Wiener model (see \cite{Wills2013}) proposed in this work is obtained by simplifying the electromechanical–aerodynamic model so that system dynamics reduce to linear first-order lags driven by the commanded inputs, while thrust retains its nonlinear dependence on 
\(\omega\) and \(\beta\) through a static output map. However, the key idea proposed in this work is that this nonlinear mapping is not fixed but refined from data in dynamic conditions, allowing deviations from the aerodynamic law. Thus, the Wiener model is not a strict reduction of the full equations, but a data-driven adaptation that preserves a simple dynamic structure while recovering part of the lost nonlinearities.

Assuming the inflow dynamics are fast (\(\tau_\lambda\) small), Eq.~\eqref{eq:lambda_dyn} is neglected and the inflow ratio is set to its quasi–steady value, linking thrust directly to the states:
\begin{equation}
\label{eq:wiener_thrust}
\hat T(t) = h\bigl(\omega(t),\beta(t)\bigr),
\end{equation}
with $h(\cdot)$ identified from dynamic data to compensate modeling simplifications. We adopt
\begin{equation}
\label{polynomial_h}
\hat T(\omega,\beta)=c_{20}\omega^2 + c_{11}\omega\beta + c_{21}\omega^2\beta + c_{12}\omega\beta^2 + c_{30}\omega^3,
\end{equation}
with coefficients $\mathbf{c}=[c_{20},c_{11},c_{21},c_{12},c_{30}]$ chosen for accuracy–parsimony balance.

For Eqs.~\eqref{eq:rotordyn}--\eqref{eq:motorelec}, electrical dynamics are taken much faster than mechanical, so \(I_m \approx ({u_\omega V_{\text{in}} - k_i \omega})/{R_m}\). Linearizing the aerodynamic torque around \((\omega_0,\beta_0)\) gives
\begin{equation}
Q_a(\omega,\beta) - Q_a(\omega_0,\beta_0) \;\approx\; C_Q^\omega\,(\omega - \omega_0)\;+\; C_Q^\beta\,(\beta - \beta_0),
\label{eq:torqlin}
\end{equation}
with \(C_Q^\omega=\partial Q_a/\partial\omega\), \(C_Q^\beta=\partial Q_a/\partial\beta\) at the operating point. Using the equilibrium condition
$k_g I_{m0} - k_\omega \omega_0 - Q_a(\omega_0,\beta_0) = 0$ yields
\[
J \,\Delta \dot{\omega} \;=\; k_g \,\Delta I_m \;-\; \bigl(k_\omega + C_Q^\omega \bigr)\,\Delta \omega \;+\; C_Q^\beta \,\Delta \beta,
\]
showing current and pitch drive \(\Delta\omega\) with effective damping \(k_\omega+C_Q^\omega\). A low-level PI on motor current uses
\[
e_\omega(t)=\Delta \omega(t)-\Delta \omega_{\text{ref}}(t),\qquad
z_\omega(t)=\int_0^t e_\omega(\tau)\,d\tau,\qquad
\Delta I_m = k_P e_\omega + k_I z_\omega,
\]
giving the closed-loop relation
\begin{equation}
\label{Delta_Omega}
\tau_\omega \,\Delta \dot{\omega} + \Delta \omega \;=\; K_P\, e_\omega(t) \;+\; K_I\, z_\omega(t) \;+\; K_\beta\, \Delta \beta,
\end{equation}
with
\[
\tau_\omega = \frac{J}{k_\omega + C_Q^\omega},\quad 
K_P = \frac{k_g k_P}{k_\omega + C_Q^\omega},\quad 
K_I = \frac{k_g k_I}{k_\omega + C_Q^\omega},\quad 
K_\beta = \frac{C_Q^\beta}{k_\omega + C_Q^\omega}.
\]

For pitch actuation, neglecting electrical dynamics in \eqref{eq:pitchmech}--\eqref{eq:pitchelec} yields
\[
\Delta\ddot{\beta} + a_\beta\,\Delta\dot{\beta} = b_\beta\,u_\beta,\qquad
a_\beta=\frac{1}{J_a}\!\left(D_a + \frac{k_a k_{ia}}{R_a}\right),\quad
b_\beta=\frac{k_a V_{\text{in}}}{J_a R_a}.
\]
With a PD controller
\[
e_\beta=\Delta \beta-\Delta \beta_{\text{ref}},\quad
\dot e_\beta=\Delta\dot{\beta}-\Delta\dot{\beta}_{\text{ref}},\quad
u_\beta=K_{P\beta}(\Delta \beta_{\text{ref}}-\Delta \beta)+K_{D\beta}(\Delta\dot{\beta}_{\text{ref}}-\Delta\dot{\beta}),
\]
the closed loop is
\begin{equation}
\Delta\ddot{\beta} + (a_\beta + b_\beta K_{D\beta})\,\Delta\dot{\beta} + b_\beta K_{P\beta}\,\Delta\beta
=
b_\beta K_{P\beta}\,\Delta \beta_{\text{ref}} + b_\beta K_{D\beta}\,\Delta\dot{\beta}_{\text{ref}}.
\label{eq:delta_beta_pd}
\end{equation}

Combining \eqref{Delta_Omega} and \eqref{eq:delta_beta_pd} in deviation form
($e_\omega=\Delta\omega-\Delta\omega_{\text{ref}}$,
$e_\beta=\Delta\beta-\Delta\beta_{\text{ref}}$,
$\dot e_\beta=\Delta\dot{\beta}-\Delta\dot{\beta}_{\text{ref}}$)
gives the linearized state–space model
\begin{equation} \label{eq:Wiener1}
    \begin{bmatrix}
        \Delta \dot{\omega} \\ \dot z_\omega \\ \Delta \dot{\beta} \\ \Delta \ddot{\beta} 
    \end{bmatrix}
    =
    \begin{bmatrix}
        \dfrac{K_P-1}{\tau_\omega} & \dfrac{K_I}{\tau_\omega} & \dfrac{K_\beta}{\tau_\omega}  & 0 \\
        1 & 0 & 0 & 0 \\
        0 & 0 & 0 & 1 \\
        0 & 0 & -\,b_\beta K_{P\beta} & -\,(a_\beta + b_\beta K_{D\beta})
    \end{bmatrix}
    \begin{bmatrix}
        \Delta \omega \\ z_\omega \\ \Delta \beta \\ \Delta \dot{\beta}
    \end{bmatrix}
    +
    \begin{bmatrix}
        -\dfrac{K_P}{\tau_\omega} & 0 & 0 \\
        -1 & 0 & 0\\
        0 & 0 & 0\\
        0 & \,b_\beta K_{P\beta} & \,b_\beta K_{D\beta}
    \end{bmatrix}
    \begin{bmatrix}
        \Delta \omega_{\text{ref}} \\ \Delta \beta_{\text{ref}} \\ \Delta \dot{\beta}_{\text{ref}}
    \end{bmatrix}.
\end{equation}

Major simplifications follow if $\Delta \dot{\beta}_{\text{ref}}\!\approx\!0$, $\Delta \dot{\omega}_{\text{ref}}\!\approx\!0$, and $K_\beta\!=\!0$, i.e., slowly varying references (or much faster than loop bandwidth) and PI rejection of pitch cross–effects in $\omega$ dynamics. Each loop then approximately decouples and the closed–loop transfer functions take the two–lag, unity–DC–gain form
\begin{align}
    G_{CL,\omega}(s)
    &= \frac{\Omega(s)}{X_\Omega(s)} \,\frac{X_\Omega(s)}{\Delta\Omega_{\text{ref}}(s)} 
    = \frac{1}{(\tau_{\omega,1}s+1)(\tau_{\omega,2}s+1)}, 
    \label{eq:GCLomega} \\
    G_{CL,\beta}(s) 
    &= \frac{B(s)}{X_B(s)} \,\frac{X_B(s)}{\Delta B_{\text{ref}}(s)} 
    = \frac{1}{(\tau_{\beta,1}s+1)(\tau_{\beta,2}s+1)},  
    \label{eq:GCLbeta}
\end{align}
with auxiliary variables $X_\Omega(s)$, $X_B(s)$ introduced by factorization and references $\Delta\Omega_{\text{ref}}(s)$, $\Delta B_{\text{ref}}(s)$ the Laplace transforms of $\Delta\omega_{\text{ref}}(t)$, $\Delta\beta_{\text{ref}}(t)$.

The effective time constants relate to controller gains and actuator parameters. For pitch (PD–closed second order):
\[
\tau_{\beta,1}\tau_{\beta,2}=\frac{1}{b_\beta K_{P\beta}},\qquad
\tau_{\beta,1}+\tau_{\beta,2}=\frac{a_\beta + b_\beta K_{D\beta}}{b_\beta K_{P\beta}}.
\]
For the RPM loop (PI–closed second order), assuming the zero lies outside the loop bandwidth and using a fast-zero approximation, the relations are
\[
\tau_{\omega,1}\tau_{\omega,2}\approx\frac{\tau_\omega}{K_I},\qquad
\tau_{\omega,1}+\tau_{\omega,2}\approx\frac{1+K_P}{K_I}.
\]

Under these assumptions, \eqref{eq:Wiener1} reduces to the fourth–order realization
\begin{align}\label{eqn4thorder}
    \begin{bmatrix}
        \Delta\dot{\omega} \\ \Delta\dot{\beta} \\ \dot{x}_\omega \\ \dot{x}_\beta
    \end{bmatrix}
    &=
    \underbrace{\begin{bmatrix}
        -\dfrac{1}{\tau_{\omega,1}} & 0 & \dfrac{1}{\tau_{\omega,1}} & 0 \\
        0 & -\dfrac{1}{\tau_{\beta,1}} & 0 & \dfrac{1}{\tau_{\beta,1}}  \\
        0 & 0 & -\dfrac{1}{\tau_{\omega,2}} & 0 \\
        0 & 0 & 0 & -\dfrac{1}{\tau_{\beta,2}}
    \end{bmatrix}}_{\mathbf{A}}
    \begin{bmatrix}
        \Delta\omega \\ \Delta\beta \\ x_\omega \\ x_\beta
    \end{bmatrix}
    +
    \underbrace{\begin{bmatrix}
        0 & 0 \\
        0 & 0 \\
        \dfrac{1}{\tau_{\omega,2}} & 0 \\
        0 & \dfrac{1}{\tau_{\beta,2}}
    \end{bmatrix}}_{\mathbf{B}}
    \begin{bmatrix}
        \Delta\omega_{\text{ref}} \\ \Delta\beta_{\text{ref}}
    \end{bmatrix},
\end{align}
so only four parameters must be identified: \(\{\tau_{\omega,1},\tau_{\omega,2}\}\) and \(\{\tau_{\beta,1},\tau_{\beta,2}\}\). In this work, these are obtained experimentally as detailed in the following section. Equations \eqref{eqn4thorder} and \eqref{eq:wiener_thrust} constitute the proposed Wiener model.

\section{Materials and methods}
\label{sec3}

The experimental setup is described in section \ref{3p1}, along with the methodology for the static characterization. Section \ref{3p3} describes the dynamic characterization and nonlinear system identification tools, while section \ref{3p4} presents the preliminary control loop implementation.

\subsection{Experimental setup and static characterization\label{3p1}}

\begin{figure}[!ht]
    \centering
    \begin{subfigure}[t]{0.49\linewidth}
        \centering
        \includegraphics[width=0.75\linewidth]{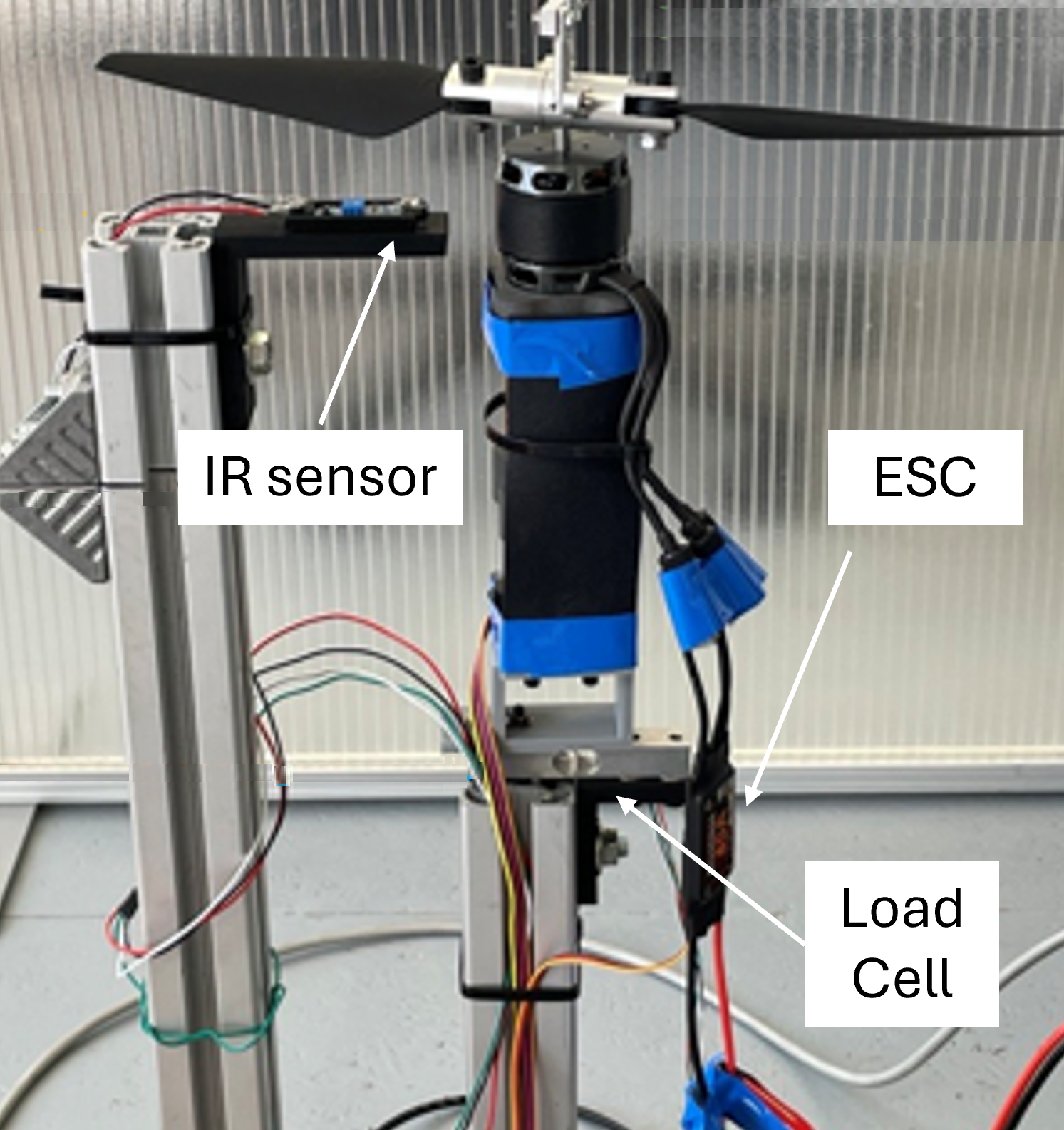}
        \caption{}
        \label{stand_photo}
    \end{subfigure}%
    \hfill
    \begin{subfigure}[t]{0.49\linewidth}
        \centering
        \includegraphics[width=0.81\linewidth]{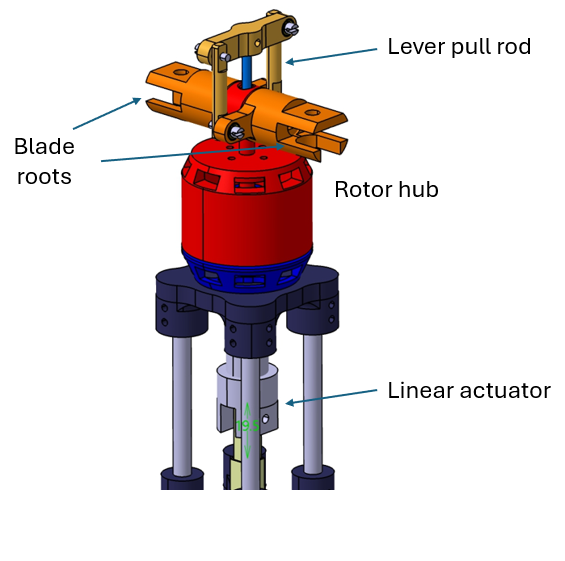}
        \caption{}
        \label{stand_mechanism}
    \end{subfigure}
    \caption{(a) Experimental setup with IR sensor, ESC, and load cell. (b) CAD model of the custom variable pitch mechanism, showing the linear actuator, rotor hub, blade roots, and lever pull rod.}
    \label{stand_photo_mechanism}
\end{figure}

A picture of the experimental setup is shown in Figure \ref{stand_photo}. The VPP mechanism is a custom-based design developed at the von Karman Institute, shown in Figure \ref{stand_mechanism}. The variable pitch mechanism consists of a hollow main shaft that houses a linear actuator. The actuator drives a vertical push–pull rod that extends through the shaft and connects to a lever system positioned above the rotor hub. The lever translates the axial motion of the actuator into a rotational motion at the blade roots, thereby adjusting the pitch angle of the blades. 

The blades were obtained by slicing a standard 1045 propeller (from root to tip) originally designed for S500/X500 platforms \citep{aerialshop_prop1045}. This procedure preserved the original aerodynamic geometry, including the airfoil cross-section, chord distribution, and twist profile. The rotor is powered by a three-phase brushless DC motor (T-Motors AT3520, KV 550) driven by a 70 A electronic speed controller (ESC). Pitch control is provided by a linear actuator (Actuonix P8-25-50-12-P) integrated within the hub. Both the motor and the actuator were supplied by dedicated power units, with the motor operated at 24 V and the actuator at 15 V (supplies: RUZIZAO 600 W and IKococater 300 W). The pitch calibration was carried out via image processing within the work by \cite{Colonval2025}. This consisted of linking the Pulse Width Modulation (PWM) signal controlling the actuator position and the pitch angle of the blades. 

Thrust was measured using a single-axis 10 kg load cell, with the signal amplified by an LM741 op-amp. Since the maximum thrust reached only 14 N, the load cell operated near its lower limit, making the signal noisy and of low resolution. Noise was mitigated with three 100 nF capacitors between the supply and amplifier and one 100 nF capacitor between the amplifier and ground, yielding sufficient signal quality. The conditioned signal was digitized by an Adafruit ADS1115 16-bit ADC interfaced with a Raspberry Pi 4, which also handled data acquisition and control. Taring was implemented in Python and performed before every measurement to ensure accuracy.

Angular velocity was measured with an HW201 infrared LED sensor and reflective tape on the motor hub, with signals processed by an Arduino Uno. Pulses generated at each passage were detected via the \texttt{attachInterrupt} function, and the time between two pulses was used to compute RPM in the main loop. The Arduino operated in slave mode, returning the encoded RPM value (first 4 bits) to the Raspberry Pi via serial USB upon request.

\begin{figure}[!ht]
    \centering
    \includegraphics[width=0.7\linewidth]{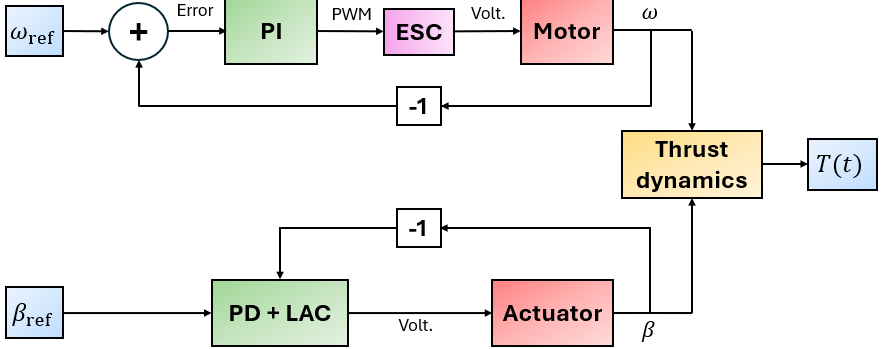}
    \caption{Block diagram for the closed-loop control architecture for motor RPM (PI controller) and blade pitch (custom PD + LAC controller).}
    \label{RPM_control}
\end{figure}

The operational PWM range for open-loop control of the BLDC motor was found to be highly sensitive to operating conditions, particularly motor heating, and prone to drift. To ensure stable performance, closed-loop feedback control of the RPM was implemented using a PI controller experimentally tuned to $K_p = 10^{-4}$ and $K_i = 2 \times 10^{-3}$. The actuator LAC board included a PD controller, tuned for fast response with gains $K_p = 10$ and $K_d = 10$. The controllers yielded satisfactory performance with the RPM PI controller effectively rejecting the disturbance caused by pitch changes. The overall control architecture is shown in Figure~\ref{RPM_control}.

The first step consisted of measuring the static thrust map, automated through a Python script. The ranges of RPM, $\omega = [2000, 6000]$, and pitch angle, $\beta = [-10, 10]$, were evenly sampled to generate a $10 \times 10$ grid of measurement points. The acquisition proceeded by iterating through the grid with two nested loops. At each point, a settling time of three seconds was allowed for transient effects to decay before data collection, and the acquisition was carried out at 250 Hz for a fixed number of samples. The thrust value was then obtained by averaging the samples over time; a sensitivity analysis indicated that 5000 samples provided a reliable estimate.

\subsection{Dynamic characterization and model fine-tuning}\label{3p3}
The experimental campaigns were conducted for 5 steps up and 5 steps down with various amplitudes and starting from 5 different initial steady states. For $\omega$, the steps up were starting from 2000 RPM and ending at [2500 - 6000] RPM. Steps down were starting from 6000 RPM and ending at [5500 - 2000] RPM. For $\beta$, the steps up were starting from -10 deg and ending at $[-6 - 10]$ deg. Steps down were starting from 10 deg and ending at $[6 - (-10)]$ deg. 

Iteration through all measurement points was again performed using two nested loops. The acquisition phase was split into two stages of five seconds. The first stage was a pre-step stage, where the transient effects from the previous step settled. The second phase was the post-step phase, where the initial estimates of parameters were extracted. Finally, data processing, normalization, and identification of characteristic time constants were performed. For both input step cases, the effective time constants in \eqref{eqn4thorder} were fitted using least squares minimization, yielding the initial estimates. By connection with the static thrust map from the static characterization, the initial Wiener model estimate is complete.


To enhance the prediction accuracy of the initial model, fine-tuning was performed. The ground-truth data were obtained by measuring an open-loop response of the thrust to $\omega_{\text{ref}}$ and $\beta_{\text{ref}}$ step sequences. To test the predictive performance for different input actuation scenarios, the open-loop input signal consisted of three parts: only $\omega_\text{ref}$ changes, only $\beta_\text{ref}$ changes, and simultaneous changes in both inputs. Model parameters were first arranged in a single vector, $\textbf{p} = \begin{bmatrix} \tau_{\omega,1} & \ \tau_{\omega,2} & \ \tau_{\beta,1} &  \tau_{\beta,2} & c_{20} & c_{11} & c_{21} & c_{12} & c_{30}\end{bmatrix}^\intercal$. The optimization algorithm aims to minimize the following cost function:
\begin{align} \label{eq_cost_FT}
    \mathcal{J}(\textbf{p}) &= \frac{1}{t_{\text{end}}}\int_0^{t_{\text{end}}} \mathcal{L}(\tau; \textbf{p}) \text{d}\tau \approx \frac{1}{N}\sum_{i=1}^N \mathcal{L}_i(\textbf{p}) \Delta t \ ,
\end{align}
where $t_{\text{end}}$ is the final time of the fine-tuning simulation, $N=t_{\text{end}}/\Delta t$ is the number of time steps, and the Lagrangian is 
\begin{align}
    \mathcal{L}_i(\textbf{p}) = \frac{1}{2}\left(\hat{T}_i(\textbf{p}) - T_i\right) ^2
\end{align}
with $\hat{T}$ and $T$ the predicted and measured thrust, respectively.

The fine-tuning was carried out using a forward sensitivity gradient descent method. This method can be utilized to update the parameters in real time. However, here it was utilized for an offline optimization. It aims to solve the problem of sensitivities, $\mathbf{s = {\partial x}/{\partial p}}$ and $\textbf{z} = {\partial \hat{T}}/{\partial \textbf{p}}$, that naturally appear in the gradient of \eqref{eq_cost_FT}, by evolving them in time as secondary state and output variables alongside the Wiener system \eqref{eqn4thorder}. The gradient of \eqref{eq_cost_FT} is computed as
\begin{align}
    \nabla_\textbf{p} \mathcal{J}(\textbf{p}) = \frac{1}{N}\sum_{i=1}^N \left(\hat{T}_i(\textbf{p}) - T_i\right)  \frac{\partial \hat{T}_i}{\partial \textbf{p}} \Delta t = \frac{1}{N}\sum_{i=1}^N \left(\hat{T}_i(\textbf{p}) - T_i\right) \textbf{z}_i \Delta t \ ,
\end{align}
and the parameters are iteratively updated via the well-known update rule $\mathbf{p = p} - \eta \nabla_\textbf{p}\mathcal{J}(\textbf{p})$, where $\eta$ is the update step. The optimization stops when the error difference between two consecutive iterations is less than a specified threshold.

\subsection{Preliminary control testing}\label{3p4}
Finally, the model-based preliminary thrust control optimization was performed. The control system comprised two PID controllers with the thrust tracking error as the input and $\omega_{\text{ref}}$ and $\beta_{\text{ref}}$ as the outputs. 

The controller equations are given as 
\begin{align}
    \omega_{\text{ref}}(t) &= K_p^\omega e_T(t) + K_i^\omega \int_0^t e_T(\tau) \text{d}\tau + K_d^\omega \frac{\text{d}e_T(t)}{\text{d}t}, \\[6pt]
    \beta_{\text{ref}}(t) &= K_p^\beta e_T(t) + K_i^\beta \int_0^t e_T(\tau) \text{d}\tau + K_d^\beta \frac{\text{d}e_T(t)}{\text{d}t}, \\[6pt]
    e_T(t) &= T_{\text{set}}(t) - T(t),
\end{align}
where $K_p^\omega, \ K_i^\omega, \ K_d^\omega$ and $K_p^\beta, \ K_i^\beta, \ K_d^\beta$ denote the proportional, integral, and derivative gains for $\omega_{\text{ref}}$ and $\beta_{\text{ref}}$, respectively, and $T_{\text{set}}$ is the thrust setpoint. 

Introducing a uniform time discretization $t\rightarrow k \Delta t$, the integral term $e_{T, \text{int}}$ was approximated using the trapezoidal rule
\begin{equation}
    e_{T, \text{int}}(t) \approx \sum_{k=1}^N \frac{1}{2}\big(e_T[k-1] + e_T[k]\big) \Delta t = e_{T, \text{int}}[k-1] + \frac{1}{2}\big(e_T[k-1] + e_T[k]\big) \Delta t, 
\end{equation} while the derivative term $e_{T, \text{der}}$ was computed using the first-order forward differences $e_{T, \text{der}}(t) \approx ({e_T[k] - e_T[k-1])/}{\Delta t}$. The clamping technique was implemented to prevent integrator windup, disabling the corresponding integrator when $\omega_{\text{ref}}$ or $\beta_{\text{ref}}$ are saturated. Since the derivative term is sensitive to the measurement noise, the first-order filter was also included:
\begin{equation}
    e_{T, {\text{der}}}[k] = (1-\delta) e_{T, {\text{der}}}[k-1] + \delta \left(\frac{e_T[k] - e_T[k-1]}{\Delta t}\right),
\end{equation} with $\delta = {\Delta t}/{T_F}$ and $T_F$ the derivative filter time constant, which was set to 0.02 s. 

\begin{figure}[!ht]
    \centering
    \includegraphics[width=1\linewidth]{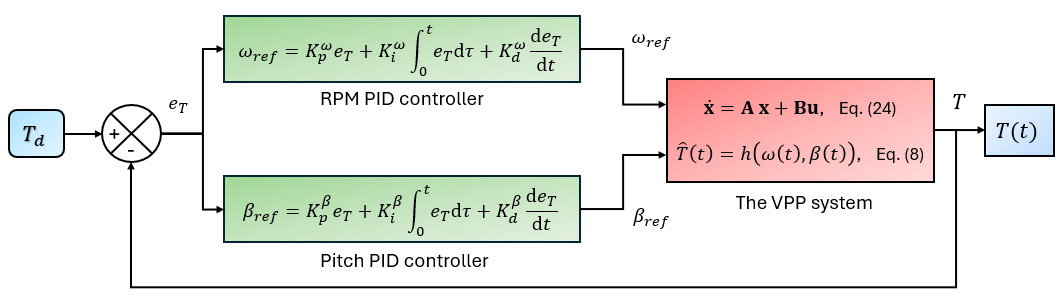}
    \caption{Block diagram of the closed-loop system.}
    \label{Closed_Loop_SS}
\end{figure}

The closed-loop system is depicted in Figure \ref{Closed_Loop_SS}. Note that it does not explicitly show the derivative filter, but the $\text{d}e_T/\text{d}t$ term is the filtered derivative. 

The controller gains were arranged in the vector $\mathbf{p_c} = \begin{bmatrix} K_p^\omega & K_i^\omega & K_d^\omega & K_p^\beta & K_i^\beta & K_d^\beta \end{bmatrix}$, and the cost function was designed as	
\begin{align}\label{cost_ctrl}
    \mathcal{J}_c(\mathbf{p_c})=\int_{0}^{t_\text{opt}}\big(T_\text{set}\left(\tau\right)-T(\tau;\mathbf{p_c})\big)^2d\tau+\alpha||\mathbf{p_c}||_2^2,
\end{align}
where $\alpha$ is the gain norm weighting coefficient, penalizing excessively large gains, and $t_\text{opt}$ is the period of the testing signal for optimization.

The optimization was carried out using the Nelder-Mead simplex algorithm \cite{Nelder1965}.

\section{Results}
\label{sec4}

\subsection{Static thrust map}
\begin{figure}[!ht] 
    \centering 
    \begin{subfigure}{0.45\textwidth}
        \centering
        \includegraphics[width=1\linewidth]{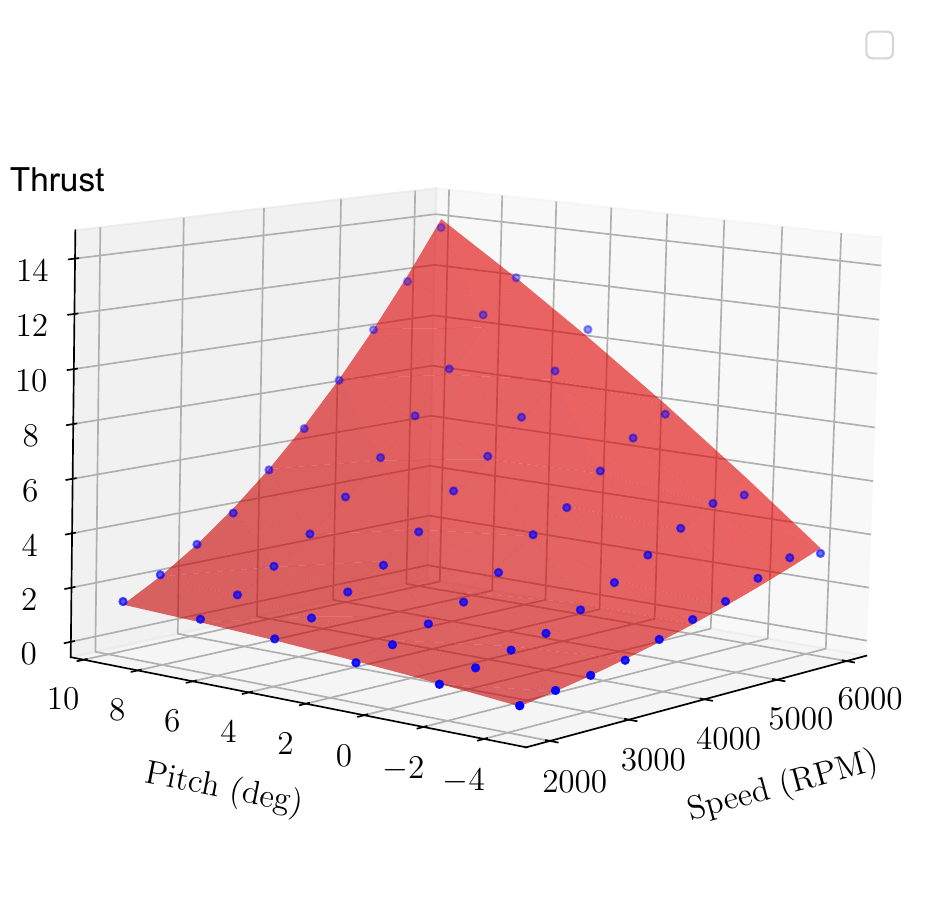}
        \caption{}
        \label{thr5000}
    \end{subfigure}
    \hspace{1mm}
    \begin{subfigure}{0.45\textwidth}
        \centering
        \includegraphics[width=1\linewidth]{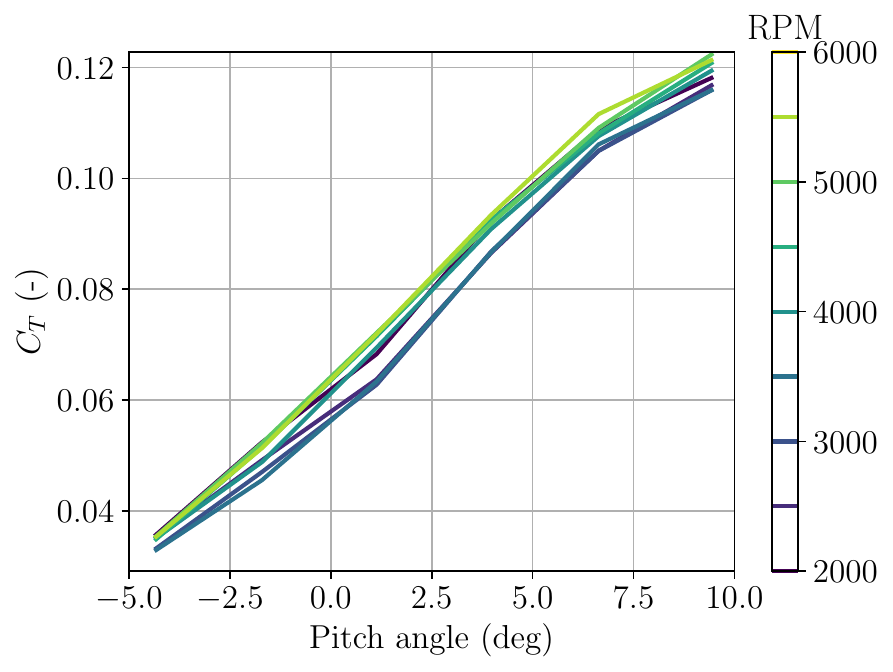}
        \caption{}
        \label{C_T}
    \end{subfigure} 
    \caption{Thrust map averaged over 5000 samples. Blue dots represent the measured data, red surface is the fitted model, $\hat{T}(\omega,\beta)$, a), and the thrust coefficient, $C_T$, b).}
    \label{thr_maps}
\end{figure}

The measured data were truncated to an operational region of the VPP with $\beta \in [-5, 10]$ deg. The measured thrust data fitted by a 3rd-order polynomial function are plotted in Figure. \ref{thr5000}. The adjusted R-squared value was $0.996$. Figure \ref{C_T} shows the associated thrust coefficient $C_T=T/(\rho n^2 D^4)$, with $n=\omega/(2\pi)$, for various  changing pitch angle. The minor impact of $\omega$ on $C_T$ suggests that the propeller is operating mostly in the linear lift regime of the blade airfoils, where thrust scales quadratically with rotation speed and geometric/inflow effects dominate over Reynolds number effects.

The fitted initial thrust map function, for $\omega, \ \beta$, and $\hat{T}$ in RPM, degrees, and Newtons, respectively, is given as
\begin{align}
    \hat{T} = h(\omega,\beta) = (0.21\omega^2 - 5.7\omega\beta + 0.023\omega^2\beta - 1.25\omega\beta^2 - 2.15\times10^{-6}\omega^3)\times10^{-6}.
\end{align}

\subsection{Dynamic model}
In the following fimures, the normalized fitted thrust responses to steps in $\omega_\text{ref}$ are shown for different $\beta = \text{const.}$ values, and responses to steps in $\beta_\text{ref}$ are shown for different $\omega = \text{const.}$ values. Initially, problems with the LAC occurred. These were subsequently solved by fine-tuning the PD controller. However, the actuator response was still problematic due to limited sensitivity to small $\beta_\text{ref}$ changes. For the small steps (1st column in both figures), the outlier responses with high noise levels were excluded.
\begin{figure}[!ht]
    \centering
    \includegraphics[width=1\linewidth]{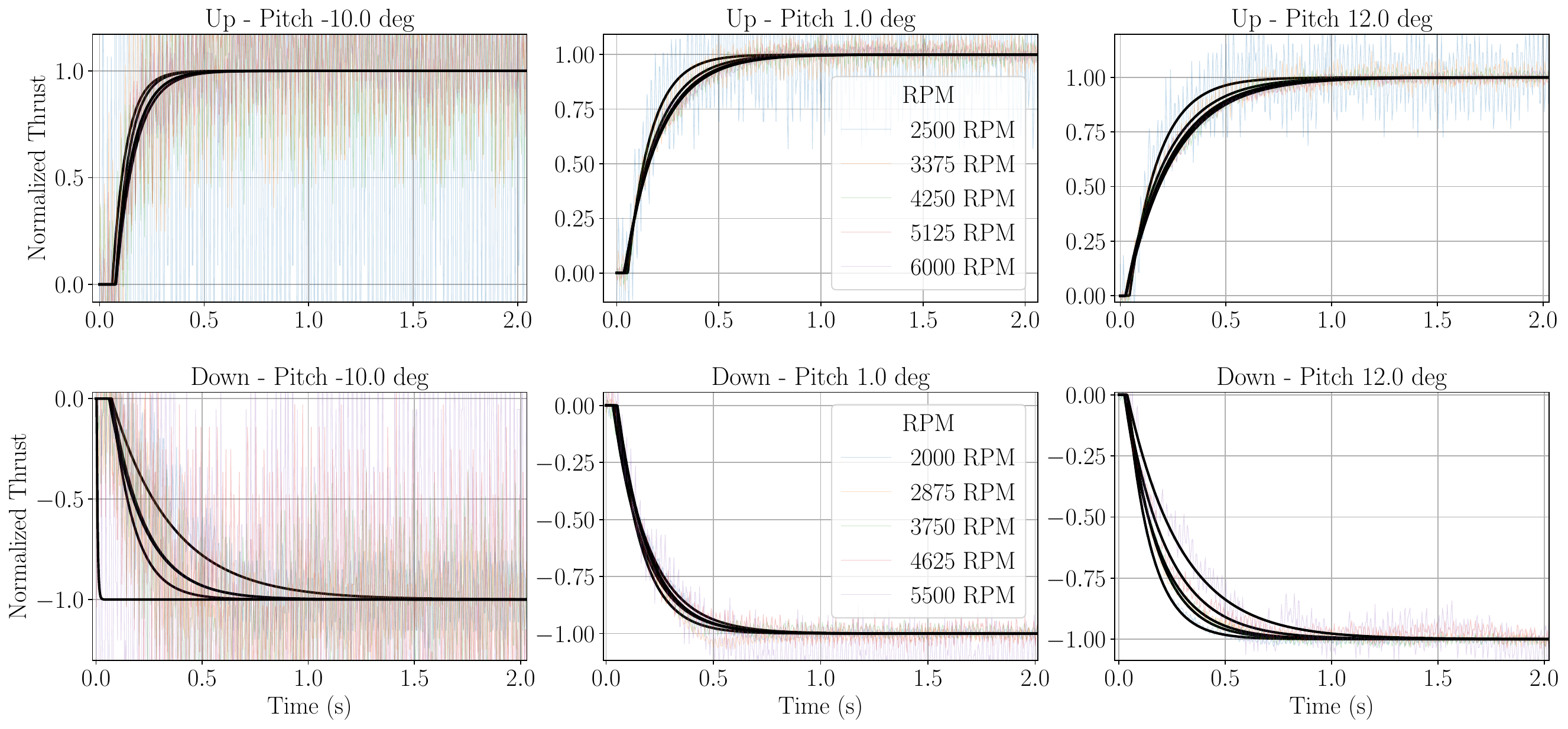}
    \caption{All fitted normalized responses to RPM steps. Colors represent measured data, and black lines are the fitted models.}
    \label{RPM_fitted_ALL}
\end{figure}
\begin{figure}[!ht]
    \centering
    \includegraphics[width=1\linewidth]{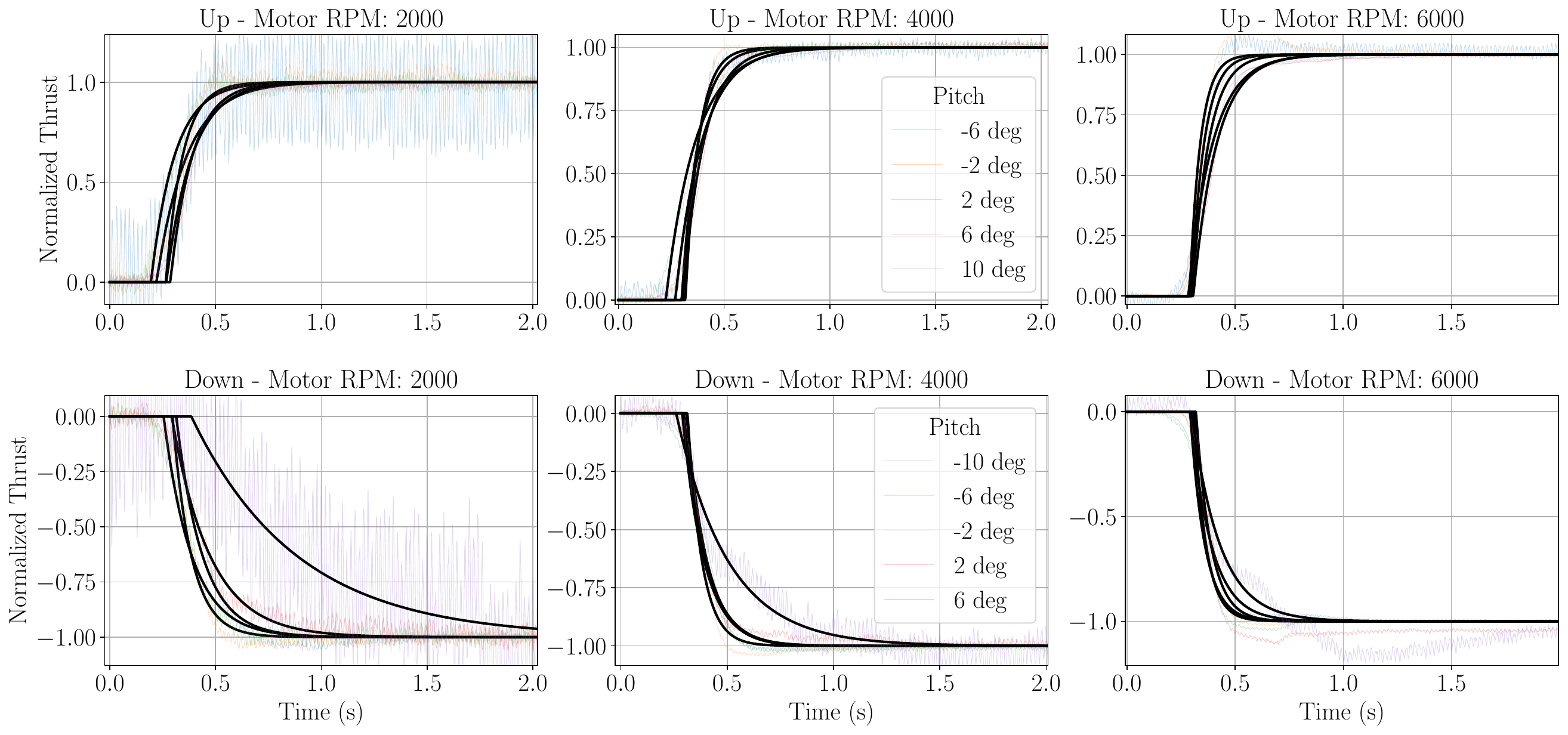}
    \caption{All fitted normalized responses to pitch steps. Colors represent measured data, and black lines are the fitted models.}
    \label{pitch_steps_all}
\end{figure}

For small thrust changes, the noise was significantly amplified by the normalization. Luckily, it did not harm the identification. There are evident overshoots for $\omega = 6000$ RPM in Figure \ref{pitch_steps_all}. The averaged time constants were $T_{\omega,1}=0.15\ s, \ T_{\beta,1}=0.11\ s, \ T_{\omega,2}=0.04\ s$, and $T_{\beta,2}=0.28\ s$. For the fine-tuning, the data were normalized as $\bar{\omega} = \omega/6000$, $\bar{\beta} = (\beta + 5)/15$, and $\bar{T} = T/15$ to preserve that $\bar{T}(0,\bar{\beta})=0$. 

\begin{figure}[!ht]
    \centering
    \includegraphics[width=0.6\linewidth]{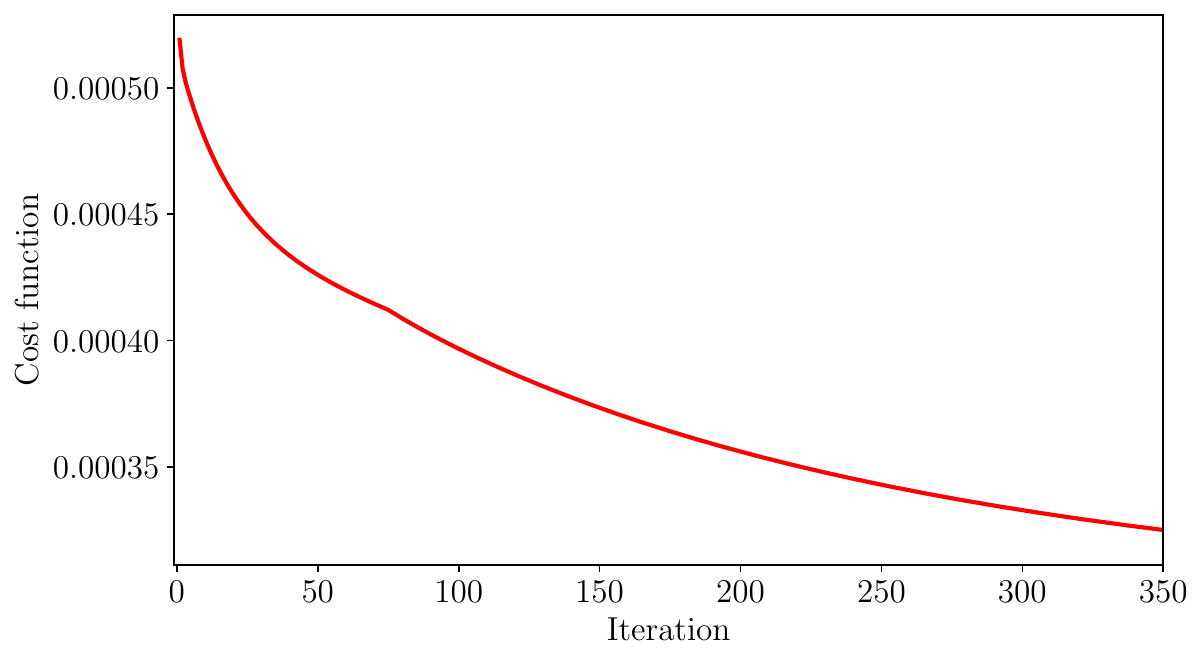}
    \caption{Fine-tuning optimization learning curve.}
    \label{sens_learn_curve}
\end{figure}

After the subsequent fine-tuning, the final model parameters were $T_{\omega,1}=0.172\ s, \ T_{\beta,1}=0.207\ s, \ T_{\omega,2}=0.053\ s$, and $T_{\beta,2}=0.374\ s$. After 350 iterations, the cost function (\ref{eq_cost_FT}) was reduced from $5.19 \times 10^{-4}$ to $ 3.2 \times 10^{-4}$, which means a $38.3\%$ accuracy increase. The time constant $T_{\beta,2}$ reflects the excessively delayed response of the pitch actuator, hampering the overall thrust response. Moreover, the model utilized the $\omega_\text{ref}$ and $\beta_\text{ref}$ signals. However, the demanded values of pitch were not always reached by the actuator, leaving, e.g., flat areas, where the model predicts bumps. This can be resolved by replacing the LAC with a custom controller, which was not realized due to time constraints. Thus, for the controller optimization, a faster time constant of $T_{\beta,2} = 0.05\ s$ was assumed. It was the lower boundary of the measured values, making the actuator response compatible with the RPM loop dynamics. The cost function evolution is depicted in Figure \ref{sens_learn_curve}.

\begin{figure}[!ht] 
    \centering 
    \begin{subfigure}{1\textwidth}
        \centering
        \includegraphics[width=0.85\linewidth]{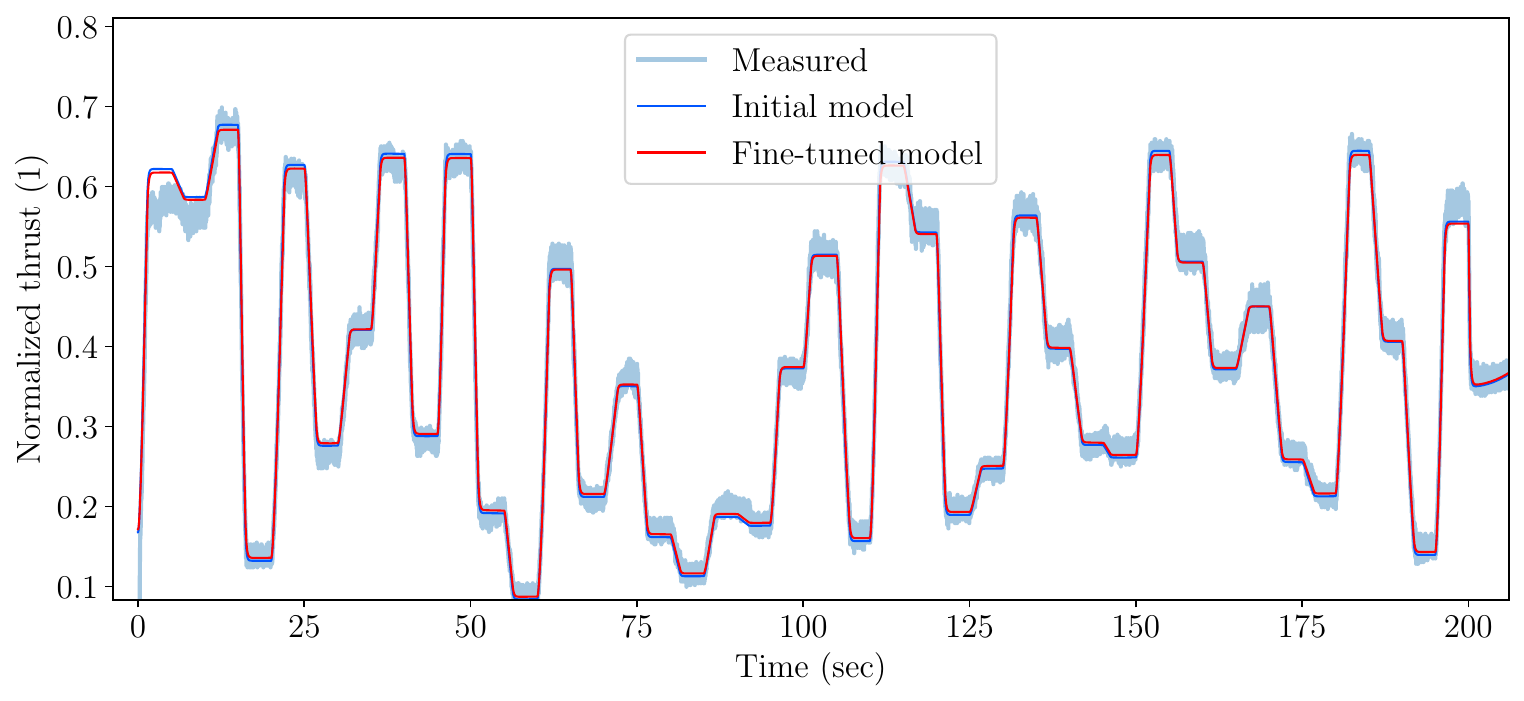}
        \caption{}
        \label{Open_Loop_RPM_Opt}
    \end{subfigure}
    \hspace{10mm}
    \begin{subfigure}{1\textwidth}
        \centering
        \includegraphics[width=0.85\linewidth]{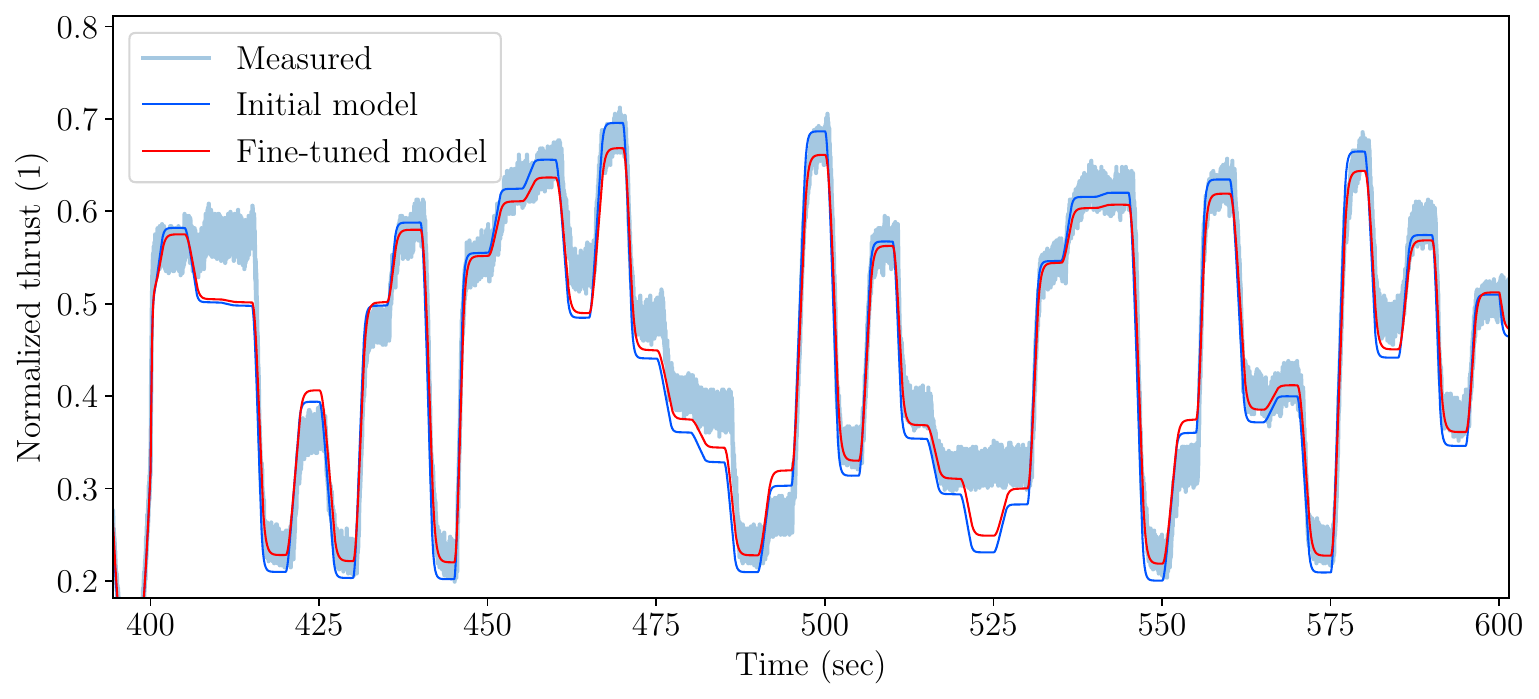}
        \caption{}
        \label{Open_Loop_Pitch_Opt}
    \end{subfigure} 
    \hspace{10mm}
    \begin{subfigure}{1\textwidth}
        \centering
        \includegraphics[width=0.85\linewidth]{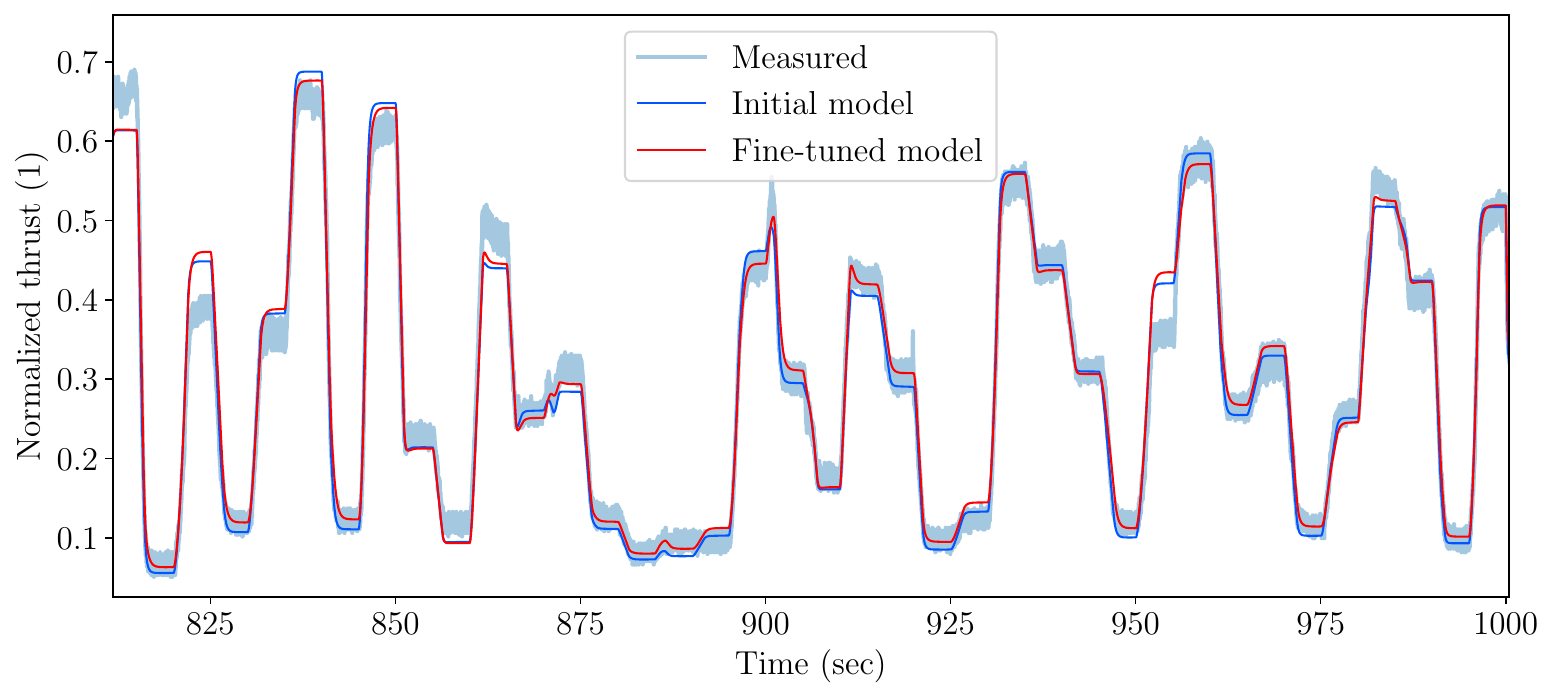}
        \caption{}
        \label{Open_Loop_BOTH_Opt}
    \end{subfigure} 
    \caption{ The comparison of the initial model (blue) and the fine-tuned model (red) for a) RPM in action, constant pitch, b) pitch in action, constant RPM, and c) both inputs in action.}
    \label{Open_Loop_Opt}
\end{figure}

The fine-tuning results are shown in Figure \ref{Open_Loop_Opt}. The improvement is most prominent at the majority of places in response to pitch only and to combined inputs.

The final thrust model is given as
\begin{align}
    \hat{\bar{T}} = h(\bar{\omega},\bar{\beta}) = 0.355\bar{\omega}^2 + 0.064\bar{\omega}\bar{\beta} + 0.8\bar{\omega}^2\bar{\beta} - 0.16\bar{\omega}\bar{\beta}^2 - 0.081\bar{\omega}^3.
\end{align}

\subsection{Control design}
The control system was designed to minimize the cost function (\ref{cost_ctrl}) for a step change of thrust setpoint, $T_\text{set}$. The goal was to prove that a combination of both inputs enhances the response compared to the optimal FPP with RPM control. Due to a high sensitivity of the system to the derivative term (even with the filter), the proportional and integral terms were optimized first, with subsequent fine-tuning of the derivative gain. All quantities were normalized before the optimization. 

\begin{figure}[!ht] 
    \centering 
    \begin{subfigure}{0.45\textwidth}
        \centering
        \includegraphics[width=1\linewidth]{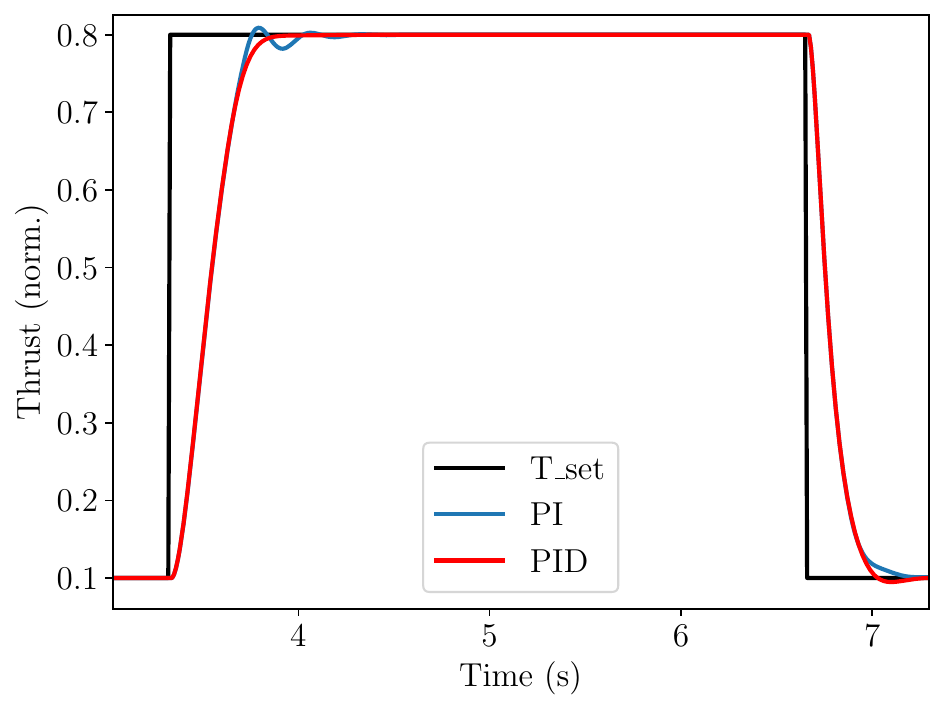}
        \caption{}
        \label{Opt_Ctrl_Thrust}
    \end{subfigure}\\
      \begin{subfigure}{0.45\textwidth}
        \centering
        \includegraphics[width=1\linewidth]{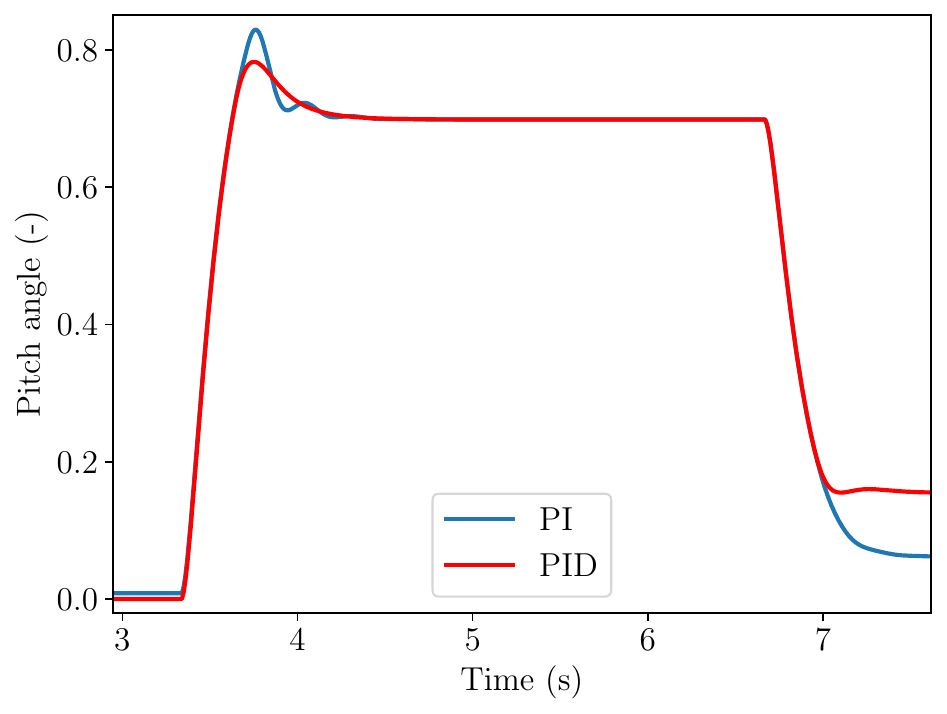}
        \caption{}
        \label{Opt_Ctrl_Beta}
    \end{subfigure} 
    \hspace{10mm}
    \begin{subfigure}{0.45\textwidth}
        \centering
        \includegraphics[width=1\linewidth]{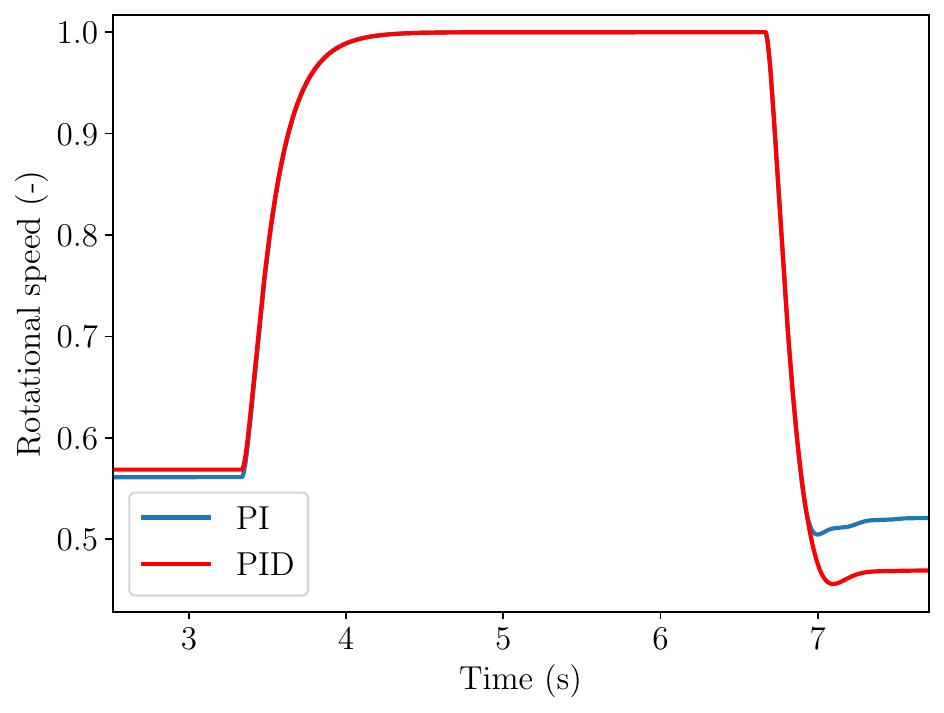}
        \caption{}
        \label{Opt_Ctrl_Omega}
    \end{subfigure} 
    \caption{Results from the step test using combined actuation via PI or PID laws. Figure a) compares the thrust step response, while figures (b) and (c) show the two actuations as a function of time.}
    \label{Opt_Ctrl}
\end{figure}

The results for the step change are depicted in Figure \ref{Opt_Ctrl}. These show that the controller effectively utilizes both control inputs to reach the optimal response without any overshoots. For the step from 0.1 to 0.8 of the normalized thrust, the setpoint was reached in approximately 0.5 s (for $t_0 = 3.33 \ s$).

\begin{figure}[!ht] 
    \centering 
    \begin{subfigure}{0.45\textwidth}
        \centering
        \includegraphics[width=1\linewidth]{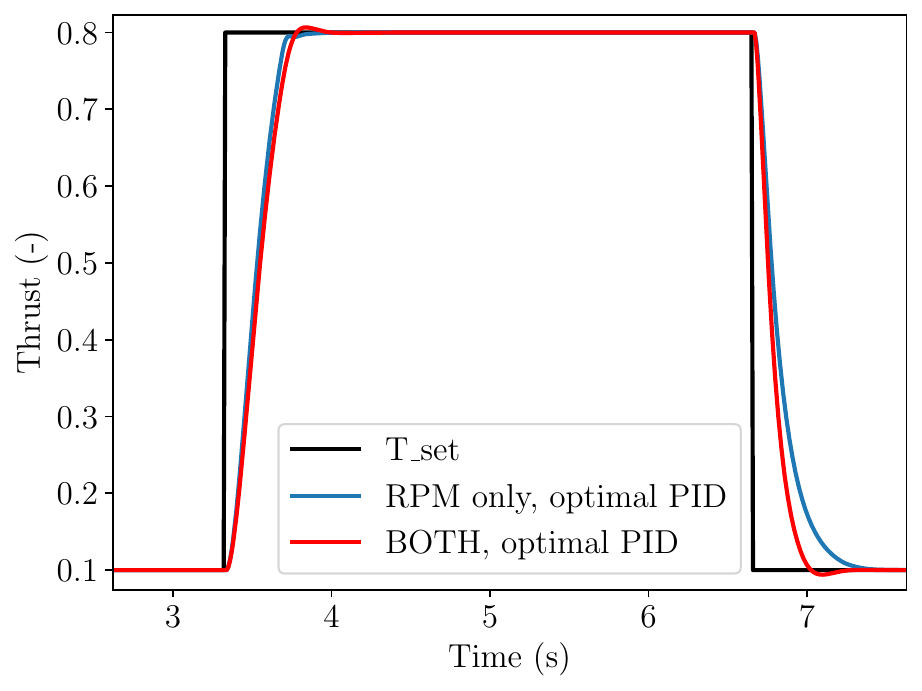}
        \caption{}
        \label{control_comparison_thrust}
    \end{subfigure}\\
    \begin{subfigure}{0.45\textwidth}
        \centering
        \includegraphics[width=1\linewidth]{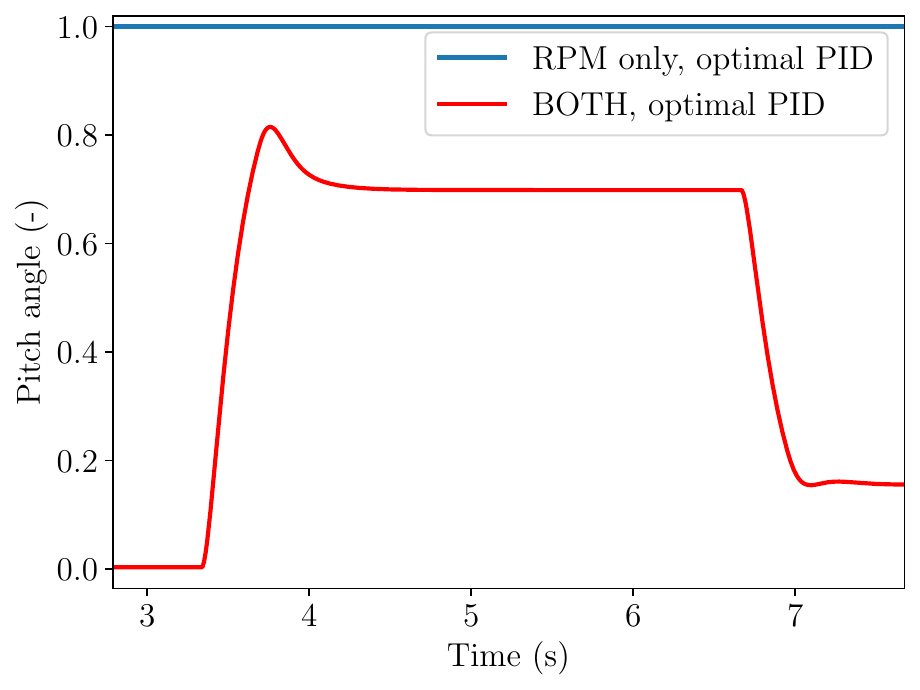}
        \caption{}
        \label{control_comparison_beta}
    \end{subfigure} 
    \hspace{10mm}
    \begin{subfigure}{0.45\textwidth}
        \centering
        \includegraphics[width=1\linewidth]{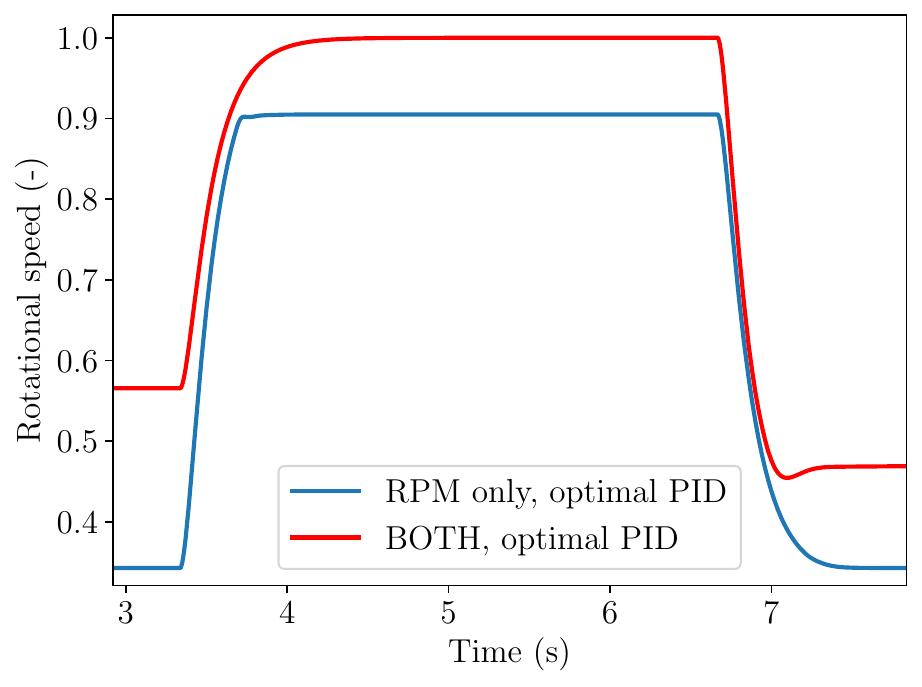}
        \caption{}
        \label{control_comparison_omega}
    \end{subfigure} 
    \caption{Step response comparison for an optimal controller solely acting on $\omega$ (blue curves) versus double actuation on both $\beta$ and $\omega$ (red curves). Figure (a) shows the thrust response, while Figures (b) and (c) show the time evolution of the two actuations.}
    \label{control_comparison}
\end{figure}

Figure \ref{control_comparison} shows the comparison of the VPP with optimal control of both inputs to the VPP with fixed pitch angle and optimal control of RPM only. The pitch was fixed at the maximum position, i.e. $\beta = 10$ deg ($\bar{\beta} = 1$). The combination of inputs exhibited a superior response, especially for the step down, when both inputs are utilized.

To summarize, the optimal gains of the RPM-only controller were $K_p^\omega = 7.47$, $K_i^\omega = 67.7$, and $K_d^\omega = 0.15$, and gains of the optimal controller using both actuations were $K_p^\omega = 9.82, \ K_i^\omega = 115.3, \ K_d^\omega = 0.318, \ K_p^\beta = 9, \ K_i^\beta = 70.44$, and $K_d^\beta = 0.25$.

\section{Conclusion}
\label{sec5}
The results show that the Wiener model is a suitable simplified model to be used as digital twin in the RT context for a variable pitch propeller. Initial parameter values can be obtained from measured thrust responses to step changes in the demanded inputs, $\omega_\text{ref}$ and $\beta_\text{ref}$, over various step amplitudes and initial steady states. Subsequent fine-tuning with a longer open-loop response time series and data assimilation improves prediction accuracy. To handle the additional delay of the LAC, it was necessary to assume similar dynamic characteristics for both actuators. By capturing open-loop thrust dynamics, the fine-tuned model includes unsteady aerodynamic effects that classical quasi-steady models neglect.


The results presented here are just preliminary. In future research, modernization of the experimental setup will be performed. In particular, it will be necessary to replace LAC with a more reliable control system, replace the IR LED RPM sensor with a high-resolution encoder, and replace the current thrust load cell with a tri-axial thrust and torque sensor with low noise levels. The identified model will be further tested in the context of reinforcement twinning.

\section*{Acknowledgements}
This work was carried out in the framework of a Short Training Program (STP) at the von Karman Institute. David Grasev is supported by the Ministry of Defence and the Ministry of Education, Youth, and Sports of the Czech Republic. This work is also part of the RE-TWIST project, which has received funding from the European Research Council (ERC) under the European Union’s Horizon Europe programme (grant agreement No 101165479). The views expressed are those of the authors and do not necessarily reflect those of the European Union or the ERC.

\bibliography{Grasev_Mendez_2026}

\end{document}